%% file: main.tex
\documentclass[sigconf, nonacm]{acmart} 
\pdfoutput=1
\AtBeginDocument{%
  \providecommand\BibTeX{{%
    \normalfont B\kern-0.5em{\scshape i\kern-0.25em b}\kern-0.8em\TeX}}}

\makeatletter
\gdef\@copyrightpermission{
  \begin{minipage}{0.2\columnwidth}
   \href{https://creativecommons.org/licenses/by/4.0/}{\includegraphics[width=0.90\textwidth]{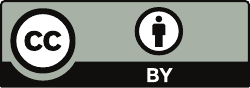}}
  \end{minipage}\hfill
  \begin{minipage}{0.8\columnwidth}
   \href{https://creativecommons.org/licenses/by/4.0/}{This work is licensed under a Creative Commons Attribution International 4.0 License.}
  \end{minipage}
}
\makeatother
\setcopyright{cc}
\setcctype[4.0]{by}
\acmConference[ScaleSys '25]{1st International Workshop on Intelligent and Scalable Systems Across the Computing Continuum}{November 19, 2025}{Vienna, Austria}
\acmBooktitle{1st International Workshop on Intelligent and Scalable Systems Across the Computing Continuum (ScaleSys '25), November 19, 2025, Vienna, Austria}

\usepackage[T1]{fontenc}
\usepackage{epstopdf}
\usepackage{graphicx}
\usepackage{placeins}
\usepackage{float}
\usepackage{nomencl}
\usepackage{balance} 
\usepackage{color,soul}
\usepackage{multirow}
\usepackage{makecell}
\usepackage{longtable}
\usepackage{hhline}
\usepackage{pifont}
\usepackage{tabularx}
\usepackage[inline,shortlabels]{enumitem}
\usepackage{geometry}
\usepackage{xspace}
\usepackage[binary-units,per-mode=symbol,exponent-product=\cdot,list-units=single,list-final-separator={,}]{siunitx}
\DeclareSIUnit[number-unit-product = ]\pixel{p}
\DeclareSIUnit[number-unit-product = ]\dB{dB}
\usepackage{chemformula}
\usepackage{gensymb}
\usepackage{amsmath}
\usepackage{booktabs}
\usepackage{comment}
\usepackage{footnote}
\usepackage{tablefootnote}
\usepackage{caption}
\usepackage{subcaption} 
\usepackage{wrapfig}
\usepackage[linesnumbered,ruled,vlined]{algorithm2e}
\usepackage[noend]{algpseudocode}

\usepackage{xurl}
\usepackage{orcidlink}
\usepackage{hyperref}
\usepackage{hyperxmp}
\usepackage{footmisc}
\hypersetup{pdfauthor=author}

\newcommand{\etal}{\emph{et al.\xspace}}

\begin{document}
\title{Predicting Encoding Energy from \\ Low-Pass Anchors for Green Video Streaming}

\author{Zoha Azimi}
\affiliation{%
  \institution{Institute of Information Technology, University of Klagenfurt}
  \city{Klagenfurt}
  \country{Austria}
}

\author{Reza Farahani}
\affiliation{%
  \institution{Institute of Information Technology, University of Klagenfurt}
  \city{Klagenfurt}
  \country{Austria}
}

\author{Vignesh V Menon}
\affiliation{%
  \institution{Video Communication and Applications Dept, Fraunhofer HHI}
  \city{Berlin}
  \country{Germany}
}

\author{Christian Timmerer}
\affiliation{%
  \institution{Institute of Information Technology, University of Klagenfurt}
  \city{Klagenfurt}
  \country{Austria}
}
\renewcommand{\shortauthors}{Zoha Azimi, Reza Farahani, Vignesh V Menon, Christian Timmerer}

\begin{abstract}
Video streaming now represents the dominant share of Internet traffic, as ever-higher-resolution content is distributed across a growing range of heterogeneous devices to sustain user Quality of Experience (QoE). However, this trend raises significant concerns about energy efficiency and carbon emissions, requiring methods to provide a trade-off between energy and QoE. 
This paper proposes a lightweight energy prediction method that estimates the energy consumption of high-resolution video encodings using reference encodings generated at lower resolutions (so-called anchors), eliminating the need for exhaustive per-segment energy measurements, a process that is infeasible at scale. We automatically select encoding parameters, such as resolution and quantization parameter (QP), to achieve substantial energy savings while maintaining perceptual quality, as measured by the Video Multimethod Fusion Assessment (VMAF), within acceptable limits. We implement and evaluate our approach with the open-source VVenC encoder on \num{100} video sequences from the Inter4K dataset across multiple encoding settings. Results show that, for an average VMAF score reduction of only \num{1.68}, which stays below the Just Noticeable Difference (JND) threshold, our method achieves \qty{51.22}{\percent} encoding energy savings and \qty{53.54}{\percent} decoding energy savings compared to a scenario with no quality degradation. 
\end{abstract}

\maketitle
\input{1_Intro}
\input{2_SotA}

\input{3_Motivation}

\input{4_Architecture}

\input{5_Setup}
\input{6_Result}

\input{7_Conclusion}
\balance
\bibliographystyle{ieeetr}
\bibliography{main}
\end{document}

%% file: 1_Intro.tex
\section{Introduction}
\label{sec:intro}
Video streaming applications, such as live content and Video-on-Demand (VoD), now dominate global Internet traffic as recent reports indicate that video content accounts for over \qty{70}{\percent} of total traffic today, with projections exceeding \qty{80}{\percent} by 2028~\cite{Sandvine_report}. HTTP Adaptive Streaming (HAS) methods such as MPEG Dynamic Adaptive Streaming over HTTP (DASH)~\cite{DASH_Survey,DASH_ref} and Apple HTTP Live Streaming (HLS)~\cite{HLS_ladder_ref} have become the de facto standard video delivery method. In these methods, each video is encoded into multiple resolution–bitrate pairs, forming a bitrate ladder, from which clients dynamically select the most suitable representation according to current network and device conditions~\cite{farahani2023sarena}. However, constructing such ladders requires encoding each video sequence into multiple representations, a process that is both computationally intensive and energy-demanding~\cite{farahani2023alive}. This energy cost is further intensified by modern video codecs such as High Efficiency Video Coding (HEVC)~\cite{HEVC} and Versatile Video Coding (VVC)~\cite{bross2021overview}, which achieve higher compression efficiency through advanced prediction, partitioning, and transform coding tools~\cite{katsenou2022energy,chachou_2023_energy}. 

This highlights a key challenge in adaptive streaming, i.e., balancing video quality, compression efficiency, and energy consumption, where the choice of the proper encoding configuration plays a pivotal role in achieving this balance~\cite{farahani2024towards,farahani2022richter}. Parameters such as resolution, framerate, and quantization parameter (QP) directly influence compression efficiency, perceptual quality, and energy consumed during encoding and decoding~\cite{farahani2024towards, qomex2024,Katsenou_vcip24}. Since higher video quality levels typically require higher energy costs, efficient configurations become essential for sustainable video streaming. While heuristic~\cite{gnostic} or Artificial Intelligence (AI)-driven methods~\cite{Rajendran_decodra,huang2021deep,menon2023transcoding} have been proposed to optimize configuration selection, they primarily focus on compression efficiency and perceptual quality, giving insufficient attention to energy considerations. Moreover, accurate energy evaluation requires per-segment energy measurements across multiple configurations, which is infeasible at scale.

This paper proposes a practical and scalable scheme that uses reference encodings generated at lower resolutions (hereafter referred to as anchors) as a proxy to predict the energy consumption of high-resolution representations. The core hypothesis is that encoding time and energy are strongly correlated, allowing patterns observed in low-resolution encodings to be leveraged for predicting the energy required at higher resolutions. For example, encoding a sequence at 360p or 540p resolution with a fixed QP typically completes much faster and consumes less energy than its 1080p or 2160p counterparts, yet still captures the content’s inherent characteristics such as motion complexity, texture richness, and scene dynamics. Using these anchors, we train machine learning (ML) models to predict higher-resolution energy consumption without exhaustive measurements, guiding an energy-aware configuration strategy that minimizes energy use while preserving perceptual quality, evaluated through Signal-to-Noise Ratio (PSNR)~\cite{psnr_ref} and Video Multimethod Fusion Assessment (VMAF)~\cite{VMAF}. The main contributions of this paper are as follows:
\begin{itemize}
\item \textbf{Dataset generation}: We construct a dataset of encoding and decoding time, energy consumption, and PSNR, VMAF scores for \num{100} video sequences from the Inter4K~\cite{inter4k_ref} dataset.
\item \textbf{Anchor-based modeling}: We introduce the concept of low-resolution anchor encodings and demonstrate that their measurements provide meaningful insights into energy consumption trends at higher resolutions. 
\item \textbf{ML-based predictions}: We develop ML models that leverage features extracted from anchor encodings to accurately estimate both energy consumption and perceptual quality.
\item \textbf{Energy-aware configuration strategy}: We design an encoding parameter selection method that uses predicted energy and quality values, minimizing energy consumption while maintaining visual quality.
\end{itemize}

%% file: 2_SotA.tex
\section{Related work}
\label{sec:sota}
\subsection{Energy consumption prediction}
Several works have proposed methods to estimate the energy consumption during video encoding or decoding. Ghasempour~\etal~\cite{ghasempour2025real} proposed a lookup table method for the fast estimation of encoding and decoding energy based on video content, resolution, and framerate features. Sharrab~\etal~\cite{sharrab2013aggregate} introduced a linear regression model for encoding energy consumption using the motion estimation range of video and QP. Azimi~\etal~\cite{qomex2024} developed Extreme Gradient Boosting (XGBoost) models to estimate encoding energy consumption using features such as video complexity, quantization parameter (QP), resolution, frame rate, and codec type. Herglotz~\etal~\cite{herglotz2015estimating} used linear regression models to estimate decoding energy based on decoding processing time. Turkkan~\etal~\cite{turkkan2022greenabr} developed a neural network-based model to predict the decoding power consumption of a video sequence using parameters such as bitrate, resolution, framerate, and file size.
Farahani~\etal~\cite{Farahani_RDIE} introduced a relative decoding energy index (RDEI), a metric that normalizes decoding energy consumption against a baseline encoding configuration, enabling cross-platform comparability and guiding energy-efficient streaming adaptations.

\textit{State-of-the-art limitations:} These methods rely on full encodings or decodings with specialized energy measurement tools, making them resource-intensive and limited in their ability to represent all encoding scenarios.
\subsection{Encoding parameter configuration}
Another line of research focuses on selecting encoding parameters to optimize video quality and efficiency. Lebreton~\etal~\cite{lebreton2023quitting} designed bitrate ladders based on user quitting probabilities, improving the perceived Quality of Experience (QoE). In parallel, ML-based approaches, such as Random Forest (RF)–based models~\cite{res_pred_ref1}, have been employed to predict optimal segment resolutions for enhanced perceptual quality. Huang~\etal~\cite{huang2021deep} proposed a reinforcement learning-based method to dynamically select bitrate-resolution pairs, jointly optimizing video quality, storage cost, and adaptation to network conditions. 
Azimi~\etal~\cite{azimi2024decoding} used XGBoost models to predict the decoding time and used a decoding time-constrained configuration setting. Rajendran~\etal~\cite{Rajendran_decodra} used Pareto-front analysis to predict
optimized framerates, constructing decoding complexity-aware ladders. 
Similarly, Katsenou~\etal~\cite{Katsenou_vcip24} used video quality, decoding time, and bitrate to optimize bitrate ladder construction.

\textit{State-of-the-art limitations:} While these approaches advance perceptual quality and compression efficiency, they largely neglect energy consumption, an increasingly critical factor for sustainable video streaming.

%% file: 3_Motivation.tex
\section{Motivation}
\label{sec:motivation}
Our approach is motivated by two key observations from preliminary experiments, enabling efficient energy prediction across multiple encoding settings.

\begin{figure}
    \centering
    \includegraphics[width=1\linewidth]{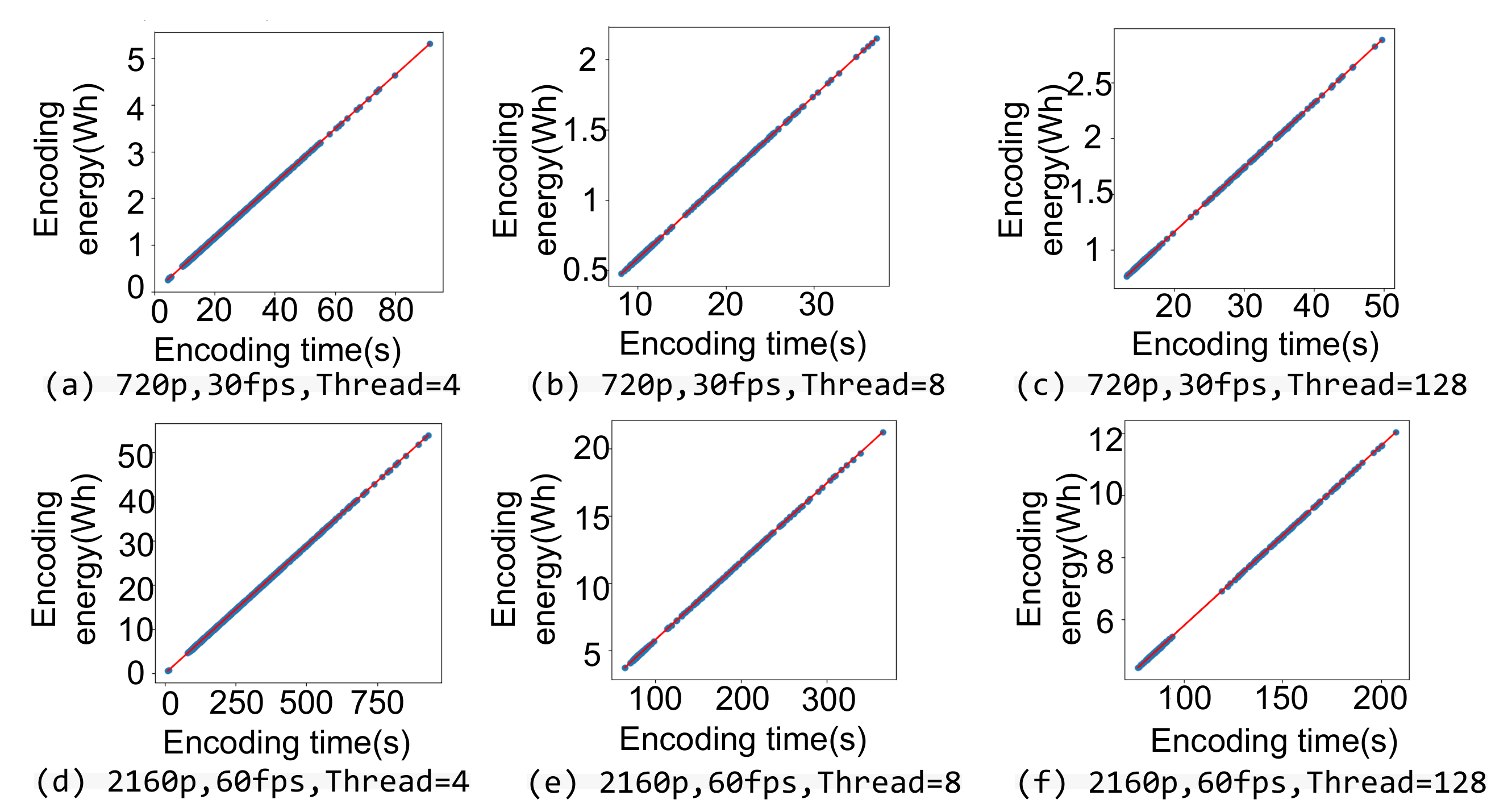}
    \caption{The correlation between encoding time and encoding energy for \num{100} video sequences, encoded with \num{720}p/\num{30}fps and \num{2160}p/\num{60}fps with three different number of threads.}
    \label{fig:intro}
\end{figure}

First, Fig.~\ref{fig:intro} shows the correlation between encoding time and energy consumption across \num{100} video sequences from the Inter4K dataset~\cite{inter4k_ref} encoded at \num{720}p/\num{30}fps and \num{2160}p/\num{60}fps. The experiments were conducted using different numbers of threads (\num{4}, \num{8}, and \num{128}). In all configurations, a strong linear correlation between encoding time and energy consumption is observed. Similar strong correlations have been observed in other works~\cite{herglotz2015estimating}. Unlike energy measurement, measuring execution time is computationally inexpensive and does not require specialized instrumentation. Thus, we use encoding time as a reliable and practical proxy for energy consumption.

\begin{figure}
    \centering
    \includegraphics[width=0.8\linewidth]{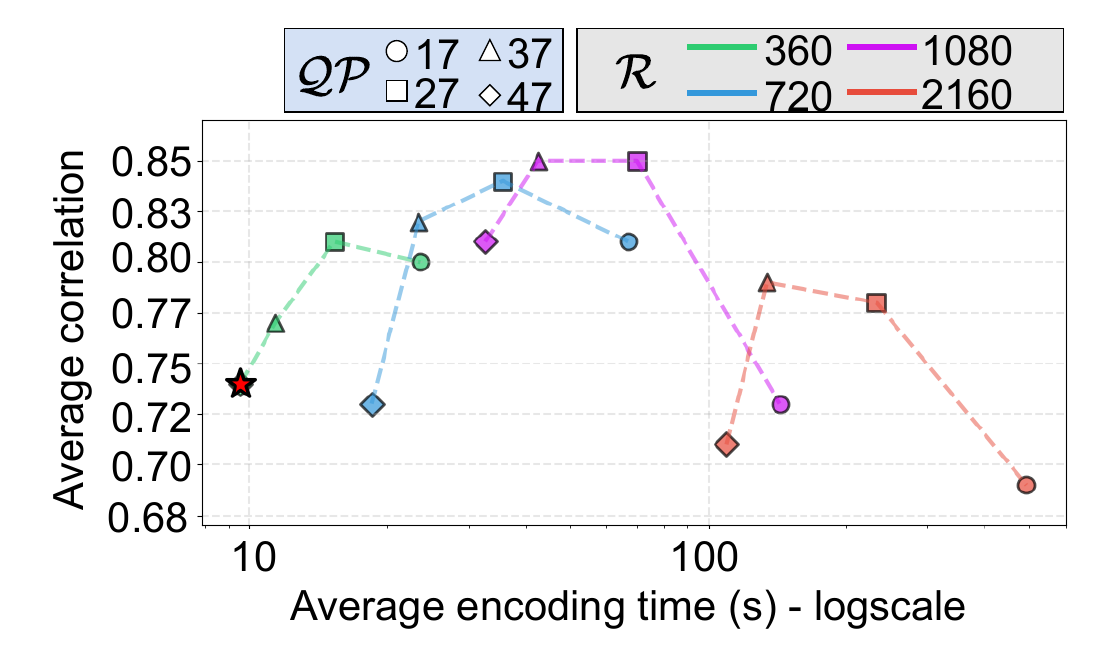}    
    \caption{Average correlation of encoding times across \num{100} video sequences for different resolutions and QPs. Each point shows the mean correlation of one configuration with all others, plotted against its average encoding time (log scale).}
    \label{fig:motiv}
\end{figure}

Second, we observe that encoding times across different representations of the same video are strongly correlated. 
Fig~\ref{fig:motiv} depicts the relationship between average encoding time (in seconds, shown on a logarithmic scale) and the average correlation of encoding times across different video representations (resolutions and QPs) across our \num{100} test video sequences. 
For each resolution-QP pair (e.g., \num{360}p, QP\num{47}), we computed the correlation between its encoding times and those of all other pairs across \num{100} video sequences. 
Overall, average pairwise correlations exceed \num{0.65}, revealing consistent temporal behavior across representations, with the correlation peaks around \num{0.8} for medium encoding settings (\num{720}p–\num{1080}p, QP:\num{27}–\num{37}). To exploit this relationship, we select the representation with the lowest resolution and highest QP as the anchor value (i.e., red star) for each video sequence, as it provides the fastest encoding with minimal computational cost, achieving a correlation of \num{0.74} while requiring only \qty{9.58}{\second} of encoding time. Thus, by measuring the anchor’s encoding time, we can infer the energy consumption of all other representations, significantly reducing computation and energy costs. This concept is extended to predict decoding energy and quality metrics using the anchor’s decoding time and perceptual quality measurements, respectively.


%% file: 4_Architecture.tex
\begin{figure*}[t]
    \centering
    \includegraphics[width=0.8\linewidth]{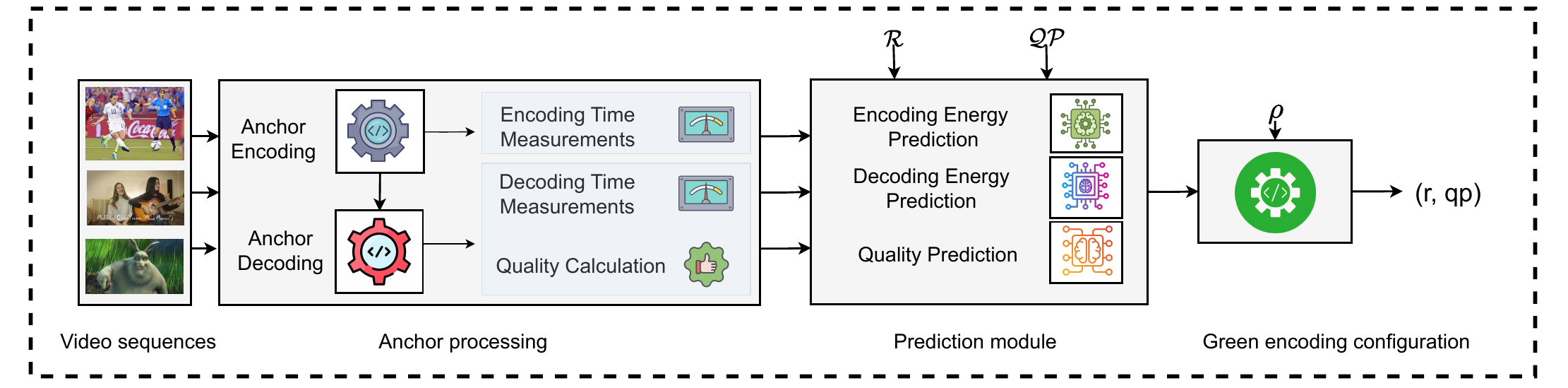}
    \caption{Proposed system architecture.}
    \label{fig:arch}
\end{figure*}

\section{System Design}
\label{sec:arch}
\subsection{Architecture}
Fig.~\ref{fig:arch} shows the proposed architecture with three main components:

\textit{(i) Anchor processing} encodes and decodes the lowest resolution ($r_{min}$) and highest quantization parameter ($qp_{max}$) representation as the anchor, measures its encoding time ($t_{enc}^A$), decoding time ($t_{dec}^A$), and quality metric ($q^A$), such as PSNR or VMAF.  

\textit{(ii) Prediction module} takes the anchor processing outputs ($t_{enc}^A$, $t_{dec}^A$, $q^A$) together with the target resolutions ($\mathcal{R}$) and QPs ($\mathcal{QP}$) as input and employs $f_{enc}$, $f_{dec}$, and $f_{q}$ to predict the encoding energy ($\hat{e}_{enc}$), decoding energy ($\hat{e}_{dec}$), and quality ($\hat{q}$) for the target video representations.
    \begin{align*}
        \hat{e}_{enc} &= f_{enc}(t_{enc}^A, r\in\mathcal{R}, qp\in\mathcal{QP}), \\
        \hat{e}_{dec} &= f_{dec}(t_{dec}^A, r\in\mathcal{R}, qp\in\mathcal{QP}), \\
        \hat{q} &= f_{q}(q^A, r\in\mathcal{R}, qp\in\mathcal{QP}). 
    \end{align*}
    
\textit{(iii) Green encoding configurations} leverages the prediction results to recommend encoding parameters ($r, qp$) based on an acceptable quality degradation factor $\rho$. When $\rho=0$, no quality degradation is allowed, and the encoding parameters are selected to provide the highest visual quality. In contrast, when $\rho=1$, the configurations prioritize minimum energy consumption, regardless of quality. 
\subsection{Execution workflow}
Algo.~\ref{alg:framework} outlines the execution workflow of our proposed method. For a given video sequence, a set of target resolutions $\mathcal{R}$ and quantization parameters $\mathcal{QP}$ are defined, along with an acceptable quality degradation factor $\rho$. First, the video sequence is encoded at the lowest resolution ($r_{min}$) and highest QP ($qp_{max}$), which serves as an anchor. The anchor’s encoding time ($t_{enc}^A$), decoding time ($t_{dec}^A$), and quality metric ($q^A$) are then measured (lines 1–2).
Next, for each resolution $r\in\mathcal{R}$ (line 4) and quantization parameter $qp\in\mathcal{QP}$ (line 5), the encoding energy prediction model $f_{enc}$ (line 6), the decoding energy prediction model $f_{dec}$ (line 7), and the quality prediction model $f_q$ (line 8) are invoked, yielding the predicted encoding energy ($\hat{e}_{enc}$), decoding energy ($\hat{e}_{dec}$), and quality ($\hat{q}$) for each representation. The predicted energies are then aggregated into $\hat{E}$ (line 9), while the quality predictions are stored in $\hat{Q}$ (line 10).
After computing the maximum obtainable quality $\hat{q}_{max}$ (line 11), all representations whose predicted quality falls within the acceptable threshold defined by $\rho$ are identified (line 12), and the one with the lowest predicted energy consumption is selected (lines 12–13). Finally, the representation $(r^*, qp^*)$ that satisfies the quality constraint and minimizes energy consumption is returned (line 14). 
The time complexity of Algo.~\ref{alg:framework} is $O(|\mathcal{R}| \times |\mathcal{QP}|)$,  where $|\mathcal{R}|$ and $|\mathcal{QP}|$ denote the number of target resolutions and quantization parameters, respectively (i.e., the number of target representations of the bitrate ladder).
\begin{algorithm}[t]
\footnotesize
\caption{Proposed Green Encoding Framework}
\label{alg:framework}
\KwIn{$\mathcal{R}$, $\mathcal{QP}$, $\rho \in [0,1]$}
\KwOut{Selected representation $(r^*, qp^*)$}

\tcp*[h]{Step 1: Anchor Processing} \\ 
$t_{enc}^A \gets Encode(r_{min}, qp_{max})$ \\
$t_{dec}^A, q^A \gets Decode(r_{min}, qp_{max})$

\tcp*[h]{Step 2: Prediction Module} \\
$\hat{E}_{enc} \gets []$, $\hat{E}_{dec} \gets []$, $\hat{Q} \gets []$ \\
\For{$r \in \mathcal{R}$}{
    \For{$qp \in \mathcal{QP}$}{
        $\hat{e}_{enc} \gets f_{enc}(t_{enc}^A, r, qp)$ \\
        $\hat{e}_{dec} \gets f_{dec}(t_{dec}^A, r, qp)$ \\
        $\hat{q} \gets f_q(q^A, r, qp)$ \\
        $\hat{E} \gets \hat{e}_{enc} + \hat{e}_{dec}$ \\
        $\hat{Q} \gets \hat{q}$
    }
}

\tcp*[h]{Step 3: Green Configuration Selection} \\
$\hat{q}_{max} \gets  Max(\hat{Q})$ \\
$\mathcal{F} \gets \{(r, qp) \mid \hat{Q} \geq (1-\rho) \cdot \hat{q}_{max}\}$ \\
$(r^*, qp^*) \gets \arg\min_{(r, qp) \in \mathcal{F}} \hat{E}$ \\
\Return $(r^*, qp^*)$
\end{algorithm}

%% file: 5_Setup.tex
\section{Evaluation Setup}
\label{sec:setup}
We conducted all experiments on a server with a \num{128}-core Intel Xeon Gold CPU and two NVIDIA Quadro GV100 GPUs. The following subsections describe dataset characteristics and analysis, ML-based prediction models, and evaluation metrics.
\subsection{Dataset analysis}
\begin{figure}
    \centering
    \includegraphics[width=0.9\linewidth]{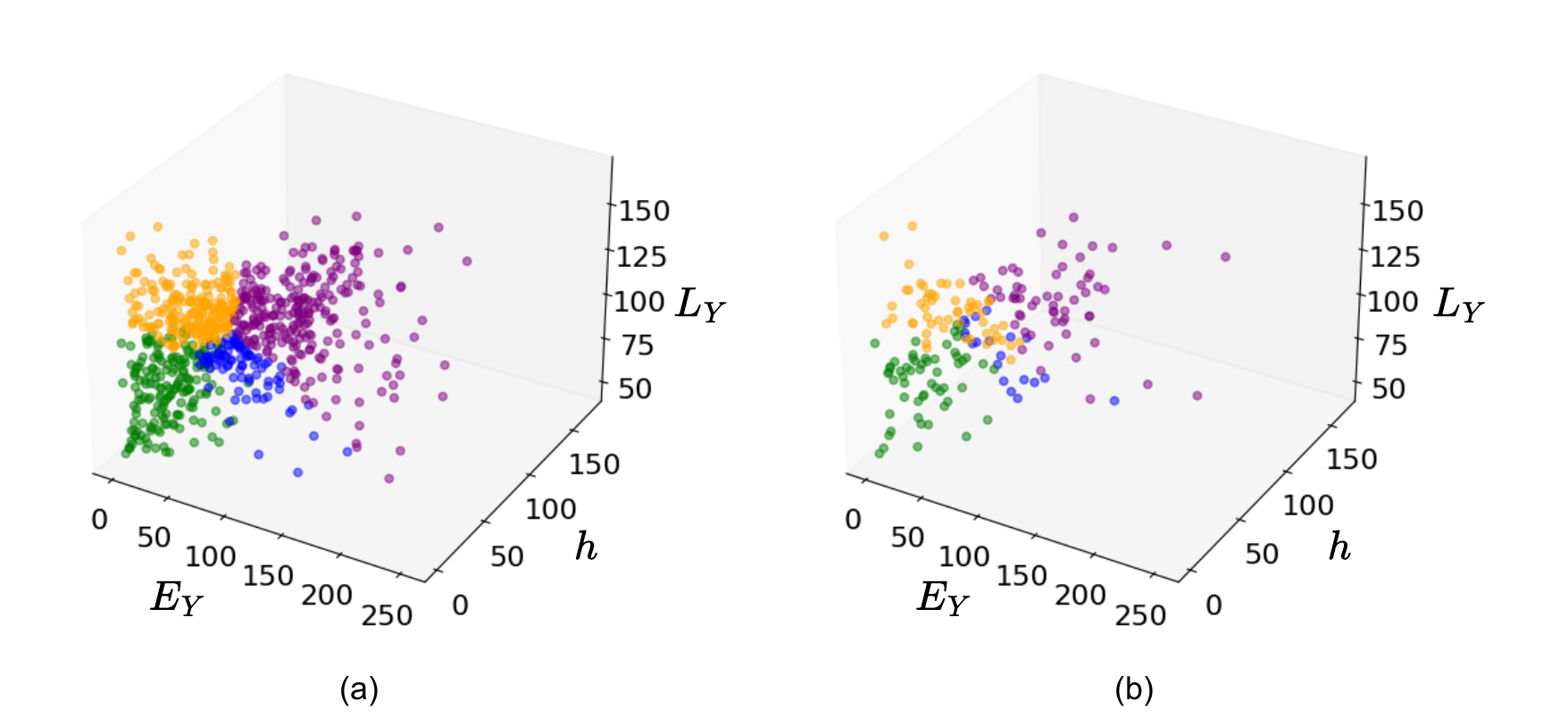}
    \caption{SOM-based clustering on the video complexity features on (a) the full dataset, and (b) our \num{100} subset.}
    \label{fig:clustering}
\end{figure}

We used \num{100} ultra-high-definition (UHD) video sequences with diverse spatiotemporal characteristics from the Inter4K dataset~\cite{inter4k_ref}. To verify that the selected subset is representative of the full Inter4K dataset, which contains \num{1000} sequences, we applied a Self-Organizing Map (SOM)~\cite{kohonen-som} clustering on the content-complexity features $(E_Y,h,L_Y)$ using the Video Complexity Analyzer (VCA) tool~\cite{vca_ref} for both the full dataset and our subset. As shown in Fig.~\ref{fig:clustering}, the selected subset spans all clusters, confirming that it is representative of the overall dataset in terms of video complexity. Table~\ref{tab:clusters} reports the statistical distribution of the content-complexity features ($E_Y$, $h$, $L_Y$) across the SOM clusters for both the full dataset and our selected subset. While minor variations exist in individual cluster values (e.g., higher $h$ and lower $E_Y$ in Cluster~\num{2} for the subset), the overall ranges and mean values remain consistent.
\begin{table}[t]
\centering
\caption{Statistical distribution of $E_Y$, $h$, and $L_Y$ across SOM clusters for the full dataset and our subset.}
\label{tab:clusters}
\fontsize{7pt}{7pt}\selectfont
\setlength{\tabcolsep}{4pt} 
\renewcommand{\arraystretch}{0.98} 
\begin{tabular}{cl|rrr|rrr|rrr}
\toprule
\textbf{Cl.} & \textbf{Set} &
\multicolumn{3}{c|}{$E_Y$} &
\multicolumn{3}{c|}{$h$} &
\multicolumn{3}{c}{$L_Y$} \\
\cmidrule(lr){3-5} \cmidrule(lr){6-8} \cmidrule(lr){9-11}
 & & Avg. & min & max & Avg. & min & max & Avg. & min & max \\
\midrule
0 & Full   & 107.4 & 37.4 & 252.4 & 59.4 & 10.4 & 171.7 & 116.3 & 59.9 & 138.9 \\
(Purple)  & Sub.   &  98.8 & 37.4 & 252.4 & 66.1 & 18.9 & 165.8 & 118.9 & 97.8 & 138.6 \\
\midrule
1 & Full   & 41.1 &  2.7 &  90.4 & 25.4 &  1.9 &  72.6 & 123.7 & 99.2 & 166.9 \\
(Orange)  & Sub.   & 48.6 &  8.8 & 113.6 & 20.5 &  2.6 &  45.3 & 125.2 & 111.0 & 166.9 \\
\midrule
2 & Full   & 87.4 & 48.2 & 174.9 & 22.1 &  2.8 &  78.8 & 100.7 & 52.9 & 122.2 \\
(Blue)  & Sub.   & 70.8 & 23.3 & 168.2 & 44.3 & 15.1 &  68.5 &  94.8 & 72.5 & 113.9 \\
\midrule
3 & Full   & 29.0 &  0.9 &  84.7 & 15.3 &  0.5 &  59.6 &  83.2 & 47.4 & 111.5 \\
(Green)  & Sub.   & 31.6 &  1.0 &  92.2 & 14.4 &  0.5 &  44.7 &  87.2 & 47.4 & 108.9 \\
\bottomrule
\end{tabular}
\end{table}

We encoded each video sequence at \num{60}~fps using \texttt{VVenC v1.11}~\cite{vvenc_ref} encoder with the \texttt{faster} preset~\cite{preset_ref}. The encoding configuration includes resolutions $\mathcal{R} = \{360, 540, 720, 1080, 1440, 2160\}$p and quantization parameters $\mathcal{QP} = \{17, 22, 27, 32, 37, 42, 47\}$~\cite{boyce_jvet_j1010_2018}. The decoding process was applied using \texttt{VVdeC v2.3.0}~\cite{VVdeC_ref}.
We recorded the encoding time, encoding energy consumption, decoding time, decoding energy consumption as well as quality scores measured by PSNR and VMAF. The energy consumption was measured with the CodeCarbon tool~\cite{codecarbon_ref}, which tracks the energy consumption of the underlying hardware using Running Average Power Limit (RAPL)~\cite{rapl} for the CPU and nvidia-ml-py~\cite{pyNVML} for the GPU. 

\begin{figure*}
    \centering
    \includegraphics[width=0.9\linewidth]{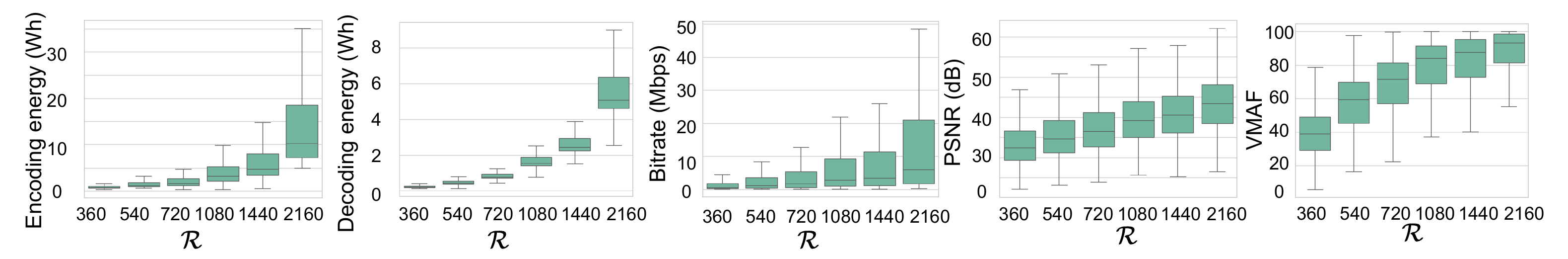}
    \caption{Impact of resolution variations on encoding and decoding energy, bitrate, PSNR, and VMAF across~\num{100} video sequences.}
    \label{fig:res}
\end{figure*}

\begin{figure*}
    \centering
    \includegraphics[width=0.9\linewidth]{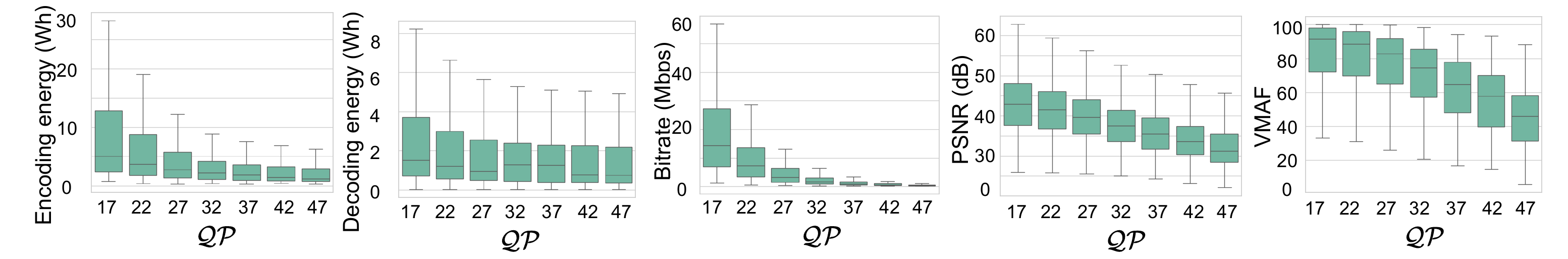}
    \caption{Impact of QP variations on encoding and decoding energy, bitrate, PSNR, and VMAF across~\num{100} video sequences.}
    \label{fig:qp}
\end{figure*}

Fig.~\ref{fig:res} presents the impact of different $\mathcal{R}$ on encoding and decoding energy, bitrate, PSNR, and VMAF. As expected, higher resolutions substantially increase both encoding and decoding energy, while improving video quality. Fig.~\ref{fig:qp} shows the variations in encoding and decoding time, bitrate, PSNR, and VMAF across different $\mathcal{QP}$ levels. As QP increases, encoding and decoding time and bitrate decrease, while quality metrics deteriorate accordingly.
\subsection{Prediction models} 
We evaluated six well-known ML models covering four different categories:
\begin{enumerate*}
\item \emph{Linear Regression} (LR)~\cite{lr_ref} and \emph{Ridge Regression} (Ridge)~\cite{mcdonald2009ridge} as linear models;
\item \emph{Random Forest} (RF)~\cite{rf_ref} as a tree-based ensemble model;
\item \emph{XGBoost} (XGB)~\cite{xgboost_ref} and \emph{LightGBM} (LGBM)~\cite{ke2017lightgbm} as gradient boosting-based ensembles;
\item \emph{Multi-Layer Perceptron} (MLP)~\cite{popescu2009multilayer}, a neural network with fully connected layers.
\end{enumerate*}

We partitioned the dataset into training (\qty{70}{\percent}) and testing (\qty{30}{\percent}) sets at the video level, ensuring that all segments from a given video belong exclusively to one set, avoiding data leakage. We randomly shuffled video identifiers with a fixed seed, guaranteeing reproducibility and consistent splitting across the three prediction tasks. We applied GridSearchCV~\cite{grid} from Scikit-learn~\cite{scikit-learn} with five-fold cross-validation on the training set.  We performed an exhaustive grid search over predefined hyperparameter spaces as summarized in Table~\ref{tab:Predictors_hyperp}. While the LR model has no tunable parameters, the Ridge model requires optimization of the regularization parameter $\alpha$. Tree-based and gradient boosting models required optimization of the number of estimators ($n_{\text{trees}}$), maximum tree depth ($d_{\text{max}}$), and learning rate ($\eta$), along with model-specific parameters such as subsampling rates or the number of leaves. For the MLP model, we fine-tuned the number ($h_{\text{num}}$) and size of hidden units ($h_{\text{size}}$) and learning rate ($\eta$). 
\begin{table}[t!]
    \centering
    \caption{Hyperparameter search space explored for ML-based prediction models.}
    \label{tab:Predictors_hyperp}
    \fontsize{7pt}{7pt}\selectfont
    \begin{tabular}{ll}
    \toprule 
    \textit{Predictive model} & \textit{Explored hyperparameters} \\ 
    \midrule
    LR & None  \\ \hline
    Ridge & $\alpha \in \{0.1, 1.0, 10.0, 100.0\}$ \\ \hline
    RF & $n_{trees} \in \{50, 100, 200\}$ , 
        $d_{max} \in \{\text{None}, 10, 20\}$ \\
       & $min\_samples\_split \in \{2, 5\}$ \\
       & $min\_samples\_leaf \in \{1, 2\}$ \\ \hline
    XGBoost & $n_{trees} \in \{50, 100, 200\}$ , $d_{max} \in \{3, 6, 9\}$ \\ 
            & $\eta \in \{0.01, 0.1, 0.2\}$ , 
             $subsample \in \{0.8, 1.0\}$ \\ \hline
    LightGBM & $n_{trees} \in \{50, 100, 200\}$ , 
              $d_{max} \in \{3, 6, 9\}$ \\
             & $\eta \in \{0.01, 0.1, 0.2\}$ ,
              $num\_leaves \in \{31, 50, 100\}$ \\ \hline
    MLP & $h_{size} \in \{64, 128, 256\}$, $h_{num} \in \{1, 2\}$    \\
        & $\eta \in \{0.001, 0.01\}$ \\ 
    \bottomrule
    \end{tabular}
\end{table}

\subsection{Evaluation metrics}
We use the following metrics to evaluate the accuracy and generalization performance of the prediction models: \textit{Coefficient of determination ($R^2$)} measures the proportion of variance in the ground truth explained by the model; a higher $R^2$ indicates better predictive accuracy.  
\textit{Mean absolute error (MAE)} measures the average relative prediction error; lower MAE indicates higher accuracy.  
\textit{Root mean squared error (RMSE)} represents the square root of the average squared prediction error, penalizing large deviations more heavily; lower RMSE indicates better performance.  
\textit{Standard deviation of absolute errors (SDAE)} measures the variability of absolute prediction errors; lower SDAE indicates more consistent predictions.  

We also assess the green encoding configuration module via:
\textit{Energy savings} quantifies the reduction in average encoding and decoding energy consumption (Wh) compared to the highest-quality scenario ($\rho=0$).  
\textit{Average quality} reports the average PSNR (dB) and VMAF scores across different encoding scenarios (varying $\rho$).  
\textit{Average quality drop} measures the reduction in PSNR (dB) and VMAF relative to the highest-quality scenario ($\rho=0$).  

%% file: 6_Result.tex
\section{Evaluation Results}
\label{sec:result}
\begin{table*}[!t]
\centering
\caption{Prediction results for encoding energy, decoding energy, PSNR, and VMAF across all models.}
\label{tab:all_results}
\fontsize{7pt}{7pt}\selectfont
\begin{tabular}{|c|cccc|cccc|cccc|cccc|}
\hline
\multirow{2}{*}{\textit{Model}} 
& \multicolumn{4}{c|}{\textbf{Encoding Energy}} 
& \multicolumn{4}{c|}{\textbf{Decoding Energy}} 
& \multicolumn{4}{c|}{\textbf{PSNR}} 
& \multicolumn{4}{c|}{\textbf{VMAF}} \\ \cline{2-17}
& $R^2 \uparrow$ & RMSE $\downarrow$ & MAE $\downarrow$ & SDAE $\downarrow$
& $R^2 \uparrow$ & RMSE $\downarrow$ & MAE $\downarrow$ & SDAE $\downarrow$
& $R^2 \uparrow$ & RMSE $\downarrow$ & MAE $\downarrow$ & SDAE $\downarrow$
& $R^2 \uparrow$ & RMSE $\downarrow$ & MAE $\downarrow$ & SDAE $\downarrow$\\ \hline
\textit{LR}    & 0.85 & 2.39 & 1.06 & 2.14 & 0.91 & 0.06 & 0.03 & 0.04 & 0.86 & 2.56 & 1.94 & 1.67 & 0.64 & 13.15 & 10.82 & 7.47 \\ \hline
\textit{Ridge} & 0.84 & 2.39 & 1.06 & 2.14 & 0.91 & 0.06 & 0.03 & 0.04 & 0.86 & 2.54 & 1.93 & 1.64 & 0.66 & 12.74 & 10.67 & 6.95 \\ \hline
\textit{RF}    & 0.89 & 1.99 & 0.79 & 1.83 & 0.95 & 0.04 & 0.02 & 0.04 & 0.91 & 1.98 & 1.40 & 1.40 & 0.89 & 7.09  & 4.97  & 5.06 \\ \hline
\textit{XGB}   & 0.90 & 1.89 & 0.70 & 1.75 & 0.95 & 0.04 & 0.02 & 0.03 & 0.91 & 2.03 & 1.40 & 1.40 & 0.90 & 6.82  & 5.08  & 4.55 \\ \hline
\textbf{\textit{LGBM}}  & 0.90 & 1.88 & 0.74 & 1.73 & \textbf{0.95} & \textbf{0.04} & \textbf{0.01} & \textbf{0.03} & 0.92 & 1.94 & 1.36 & 1.37 & 0.90 & 6.77  & 5.04  & 4.52 \\ \hline
\textbf{\textit{MLP}}   
& \textbf{0.91} & \textbf{1.82} & \textbf{0.71} & \textbf{1.67} 
& 0.95 & 0.04 & 0.02 & 0.04 
& \textbf{0.92} & \textbf{1.93} & \textbf{1.33} & \textbf{1.40} 
& \textbf{0.92} & \textbf{6.40} & \textbf{4.78} & \textbf{4.26} \\ \hline
\end{tabular}
\end{table*}
\subsection{Anchor selection analysis}\label{anchor_selection}
\begin{figure}[t]
    \centering
    \includegraphics[width=0.7\linewidth]{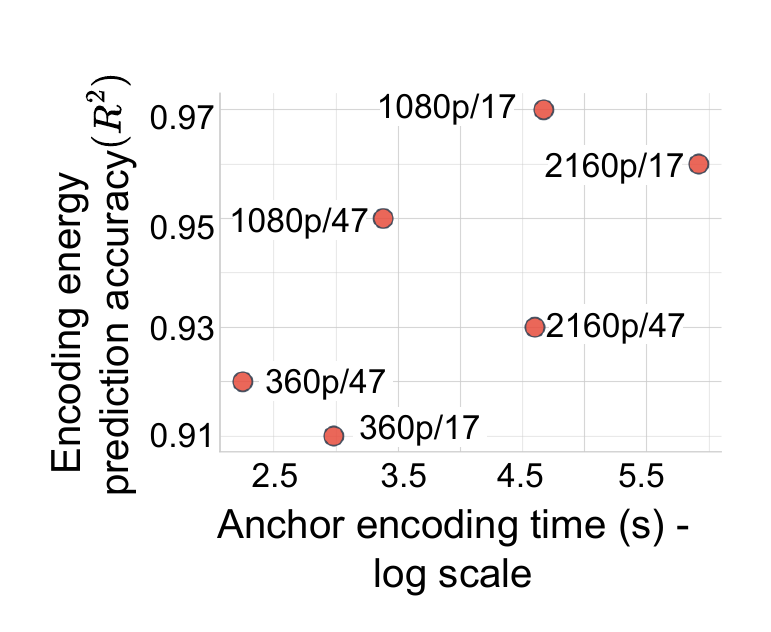}
    \caption{Encoding energy prediction accuracy using different anchors.}
    \label{fig:diff_anchor}
\end{figure}

Fig.~\ref{fig:diff_anchor} shows the accuracy of energy prediction models, when a different resolution-QP pair was selected as an anchor. We evaluated six resolution–QP pairs to cover low (360p), medium (1080p), and high (2160p) resolutions, with the minimum (QP = 17) and maximum (QP = 47). The x-axis shows the average encoding time required for each anchor, averaged across \num{100} video sequences. The results show that higher $R^2$ values are obtained at the expense of substantially longer encoding times (up to \qty{372.48}{\second} for 2160p/17), whereas the fastest configuration (\qty{9.48}{\second} for 360p/47) still achieves reasonable accuracy ($R^2 = 0.92$). These findings validate the rationale discussed in Section~\ref{sec:motivation}, confirming that the optimal anchor corresponds to the configuration with the lowest encoding time.

\subsection{Prediction models analysis}\label{prediction-results}
We evaluated the performance of the candidate prediction models on four target metrics: encoding energy, decoding energy, PSNR, and VMAF. The reported results represent the average values across the entire test set. 
\begin{enumerate}
\item{Encoding energy prediction.} 
Table~\ref{tab:all_results} (Encoding Energy) shows that MLP achieved the highest $R^2$ (\num{0.91}) and the lowest RMSE among all models, with $h_{num}=1, h_{size}=64$, and $\eta = 0.01$.
\item{Decoding energy prediction} 
Table~\ref{tab:all_results} (Decoding Energy) shows that RF, XGB, LGBM, and MLP have similar performance with $R^2$ (\num{0.95}). We selected LGBM due to its slightly lower MAE (\num{0.01}). LGBM achieved its best performance with $n_{leaves} = 50$, $n_{estimators} = 50$, $d_{\max} = 3$ and $\eta = 0.1$.  
\item{PSNR prediction} Table~\ref{tab:all_results} (PSNR) shows that MLP and LGBM achieved the best overall performance, with the highest $R^2$ (\num{0.92}) among all models. However, with slightly lower RMSE (\num{1.93}) and MAE (\num{1.33}), MLP with $h_{num}=2, h_{size} = 256, 128$ and $\eta = 0.001$ was selected for PSNR prediction. 
\item{VMAF prediction} Table~\ref{tab:all_results} (VMAF) shows that MLP achieved the best overall performance, achieving the lowest RMSE (\num{6.4}) and MAE (\num{4.78}) and highest $R^2$ (\num{0.92}) among all models with $h_{num}=2, h_{size} = 256, 128$ and $\eta = 0.001$ and therefore was selected for VMAF prediction.
\end{enumerate}
\subsection{Green encoding configuration}
Table~\ref{tab:config} reports the average energy savings (encoding and decoding) and quality degradation (PSNR, VMAF) across the test video sequences for different $\rho$ values, using the configuration with $\rho=0$ (no quality degradation) as the reference. We also report the average quality scores (PSNR and VMAF) corresponding to the selected encoding configurations.

For $\rho = 0.05$, i.e., allowing a \qty{5}{\percent} degradation in VMAF, the average quality loss remains minimal (\num{1.68} VMAF points and \qty{3.67}{\dB} PSNR) while achieving substantial energy saving of~\qty{51.22}{\percent} in encoding and \qty{53.54}{\percent} in decoding. 
The minimum noticeable quality difference, referred to as the just-noticeable-difference (JND) for VMAF, has been reported to range between \num{2} and \num{6} in prior works~\cite{amirpour2022between, kah2021fundamental, ozer2017finding}, indicating that the observed quality loss is likely imperceptible.
As $\rho$ increases, the potential energy savings grow further but at the expense of more noticeable quality degradation. For instance, at $\rho = 0.3$, energy savings exceed \qty{90}{\percent}, whereas perceived quality declines by more than \num{25} VMAF points. These results indicate that a modest quality relaxation (e.g., $\rho = 0.05$) yields substantial energy savings while having a negligible impact on perceived visual quality. 

\begin{table}[t]
    \centering
    \fontsize{7}{7}\selectfont
    \caption{Average VMAF and PSNR drops, and encoding, decoding energy savings for different $\rho$ values.}
    \label{tab:config}
        \begin{tabular}{|c|c|c|c|c|c|c|}
        \hline
        \multirow{2}{*}{$\rho$} & Avg.  & Avg.  & VMAF  & PSNR  & Enc. energy  & Dec. energy  \\ 
         & VMAF $\uparrow$ & PSNR $\uparrow$ & drop $\downarrow$ & drop $\downarrow$ & savings\% $\uparrow$ & savings\% $\uparrow$ \\ \hline
        0            & 99.74        & 50.05     & 0                 & 0              & 0                     & 0                     \\ \hline
        0.05         & 98.06        & 46.38     & 1.68              & 3.67           & 51.22                 & 53.54                 \\ \hline
        0.1          & 94.74        & 44.16     & 5                 & 5.89           & 70.75                 & 70.28                 \\ \hline
        0.3          & 72.97        & 37.57     & 26.77             & 12.48          & 92.64                 & 89.90                 \\ \hline
        0.5          & 53.06        & 33.97     & 46.68             & 16.07          & 95.82             & 94.25                 \\ \hline
        0.7          & 34.83        & 31.30     & 64.91             & 18.74          & 97.32                 & 96.88                 \\ \hline
        1            & 31.52        & 30.85     & 68.22             & 19.19          & 97.58                 & 97.42                 \\ \hline
        \end{tabular}
\end{table}

\begin{figure}
    \centering
    \includegraphics[width=1\linewidth]{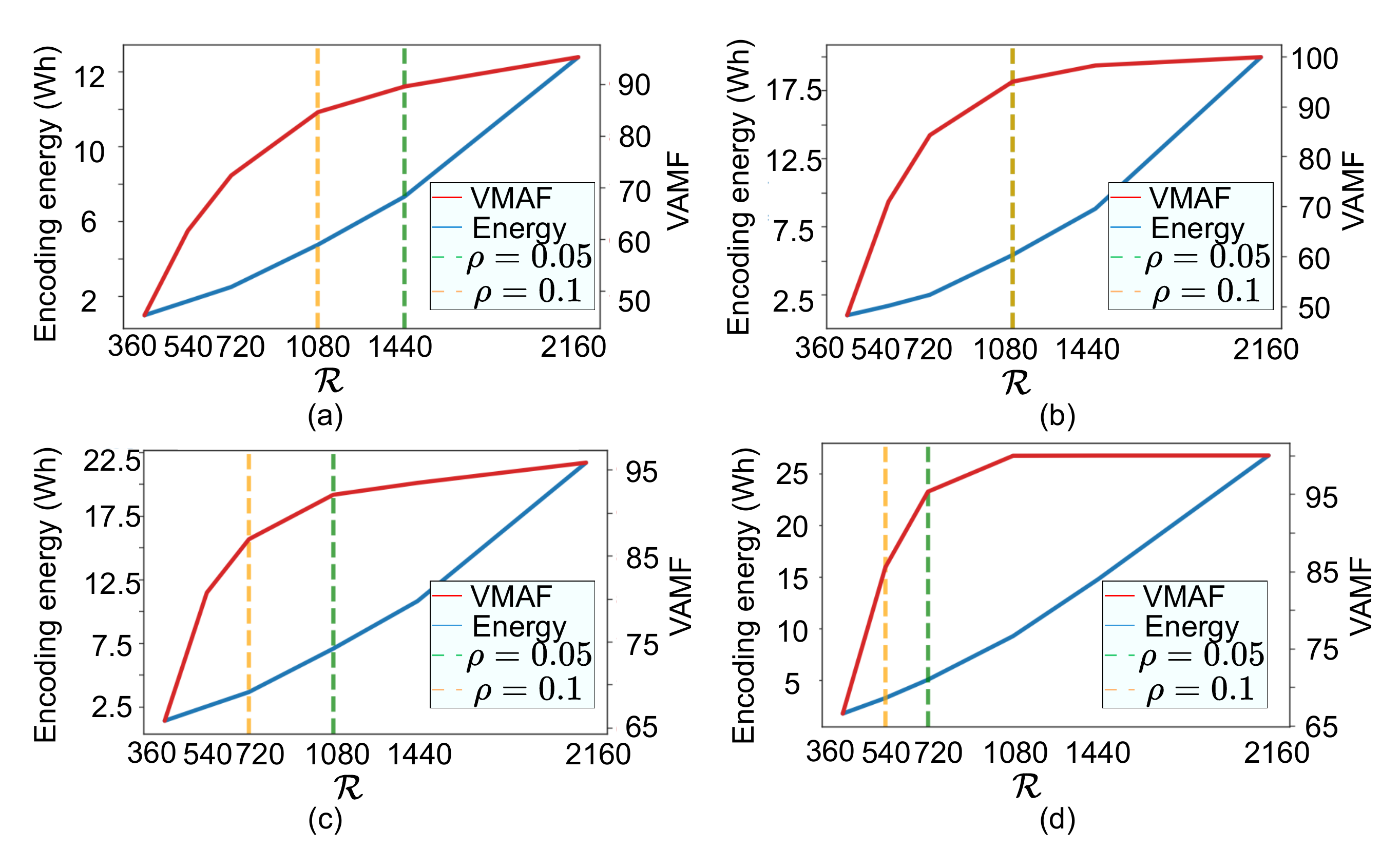}
    \caption{Impact of changing $\rho$ values on  video sequences: (a) \texttt{sequence 72} $(E_Y=73.07, h=61.2, L_Y=122.84; QP=27)$, (b) \texttt{sequence 65} $(41.61, 9.29, 90.10;17)$, (c) \texttt{sequence 23} $(25.97, 28.37, 91.22;22)$,(d) \texttt{sequence 98} $(37.39, 96.16, 109.35;17)$.}
    \label{fig:per_video}
\end{figure}

Fig.~\ref{fig:per_video} illustrates the impact of $\rho = 0.05, 0.1$ values on four video sequences with different complexity levels. Each plot shows the changes in encoding energy consumption and VMAF across resolutions. The QP is fixed to allow visualization in a two-dimensional plane. The selected resolution differs for each sequence due to variations in content complexity, such as color dynamics, scene change frequency, motion intensity, and brightness. 
For example, under $\rho=0.05$, sequence \texttt{72} (Fig.~\ref{fig:per_video} (a)) at \num{1440}p resolution achieves a \qty{50.4}{\percent} energy saving while maintaining acceptable quality compared to \num{2160}p. 
Sequence \texttt{65} (Fig.~\ref{fig:per_video} (b)) stays within \qty{5}{\percent} quality degradation at \num{1080}p. Even when the degradation threshold is increased to \qty{10}{\percent} ($\rho=0.1$), the selected resolution remains \num{1080}p, as lower resolutions would exceed the allowed quality loss.
For sequence \texttt{23} (Fig.~\ref{fig:per_video} (c)), the selected resolutions are \num{1080}p and \num{720}p for $\rho=0.05$ and $0.1$, respectively. In contrast, for sequence \texttt{98} (Fig.~\ref{fig:per_video} (d)), a lower resolution of \num{720}p is sufficient to meet the quality constraint of $\rho=0.05$, resulting in an \qty{81.1}{\percent} energy saving.

%% file: 7_Conclusion.tex
\section{Conclusions}
\label{sec:conclusion}
This paper presents a novel method for predicting energy consumption in the context of selecting configurations for green encoding by employing low-pass anchors. The proposed method involves assessing the encoding and decoding durations of the anchor encodings that are created at reduced resolutions and utilizes lightweight machine learning models to anticipate the energy usage and video quality across all other representations within a given sequence. By analyzing these predictions, the encoding configuration is meticulously chosen to strike a balance between energy efficiency and acceptable quality reduction. Our evaluation, conducted on a dataset comprising \num{100} video sequences from the Inter4K dataset, illustrates that restricting the decrease in the VMAF score to \qty{5}{\percent} results in substantial energy savings, specifically \qty{51.22}{\percent} in encoding energy and \qty{53.54}{\percent} in decoding energy, when compared to an encoding configuration selection that prioritizes maximum quality. 
